\documentclass[journal]{IEEEtran}
\usepackage[usenames]{color}
\usepackage{epsfig}
\usepackage{graphics}
\usepackage{amsmath}
\usepackage{amssymb}
\usepackage{multirow}
\usepackage{cite}
\usepackage{array}
\usepackage{pslatex} 
\usepackage{url}
\usepackage{caption}
\usepackage{booktabs}
\usepackage{lineno}
\usepackage{graphicx}  
\usepackage{setspace}
\usepackage{tikz}
\usepackage{hhline}
\usepackage{mathtools}
\usepackage[letterpaper]{geometry}
\ifCLASSINFOpdf
\else
\fi
\ifCLASSOPTIONcompsoc
\usepackage[caption=false,font=normalsize,labelfont=sf,textfont=sf]{subfig}
\else
  \usepackage[caption=false,font=footnotesize]{subfig}
\fi
\usepackage[english]{babel}
\usepackage[utf8]{inputenc}
\usepackage{fancyhdr}
\usepackage{bm}
\begin{document}
%
\title{\textcolor{violet}{Graph Attention Networks for Detecting Epilepsy from EEG Signals Using Accessible Hardware in Low-Resource Settings\vspace{0.25cm}}}

%
%
%

\author{Szymon Mazurek, 
        Stephen Moore, and Alessandro Crimi* 
\thanks{
	\par Szymon Mazurek and Alessandro Crimi are with AGH University of Krakow, al. Adama Mickiewicza 30, 30-059 Kraków, Poland. Stephen Moore is with the University of Cape Coast, Cape Coast, Ghana.  (correspondence e-mail: alecrimi@agh.edu.pl).
}}

\maketitle\thispagestyle{fancy}

\begin{abstract}
\textit{Goal:} Epilepsy remains under-diagnosed in low-income countries due to scarce neurologists and costly diagnostic tools. We propose a graph-based deep learning framework to detect epilepsy from low-cost Electroencephalography (EEG) hardware, tested on recordings from Nigeria and Guinea-Bissau. Our focus is on fair, accessible automatic assessment and explainability to shed light on epilepsy biomarkers.   \textit{Methods:} We model EEG signals as spatio-temporal graphs, classify them, and identify interchannel relationships and temporal dynamics using graph attention networks (GAT). To emphasize connectivity biomarkers, we adapt the inherently node-focused GAT to analyze edges. We also designed signal preprocessing for low-fidelity recordings and a lightweight GAT architecture trained on Google Colab and deployed on RaspberryPi devices. \textit{Results:} The approach achieves promising classification performance, outperforming a standard classifier based on random forest and graph convolutional networks in terms of accuracy and robustness over multiple sessions, but also highlighting specific connections in the fronto-temporal region. \textit{Conclusions}: The results highlight the potential of GATs to provide insightful and scalable diagnostic support for epilepsy in underserved regions, paving the way for affordable and accessible neurodiagnostic tools.
\end{abstract}


%
\IEEEpeerreviewmaketitle

\textbf{\textit{Impact Statement-} This work demonstrates that low-cost EEG combined with advanced graph-based AI can improve access to epilepsy diagnosis in low- and middle-income countries, addressing an important healthcare gap.}\\
\\

\begin{figure*}[!]
  \centering
  \includegraphics[width=0.99 \linewidth]{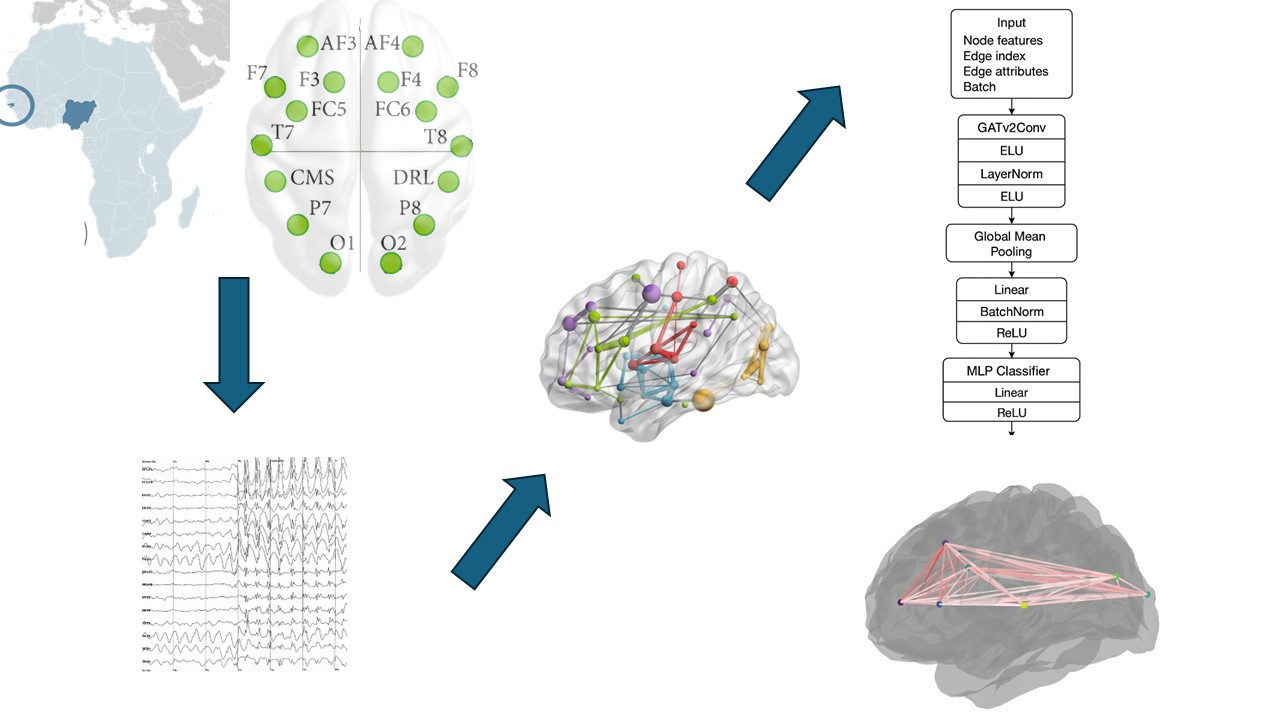}
  \caption{Overall Pipeline: Data are collected with an affordable portable EEG in low-income countries (left), they are preprocessed and  used to create brain connectivity matrices (center). Then, classified using graph attention networks, and the used weights are investigated to shed some lights about epilepsy biomarkers (right). In this figure the electrodes labels for the Epoc EEG device and the used GAT architecture are also shown.}
  \label{fig:pipeline}
\end{figure*} 

\section{Introduction}
Epilepsy is a neurological disorder characterized by recurrent seizures that result from abnormal electrical activity in the brain. It is the fourth most common neurological disorder in the world, affecting approximately 50 to 60 million people worldwide \cite{PeruccaEpilepsyWorldwide, ThurmanEpilepsyWorldwide}. In epilepsy diagnosis, electroencephalography (EEG) plays a crucial role. Currently, it is the main tool for studying the epileptogenic cortex \cite{JehiEpilepticZone}, as it can detect abnormal brain activity associated with seizures. 
Crucially, diagnosis relies on the correct identification of specific patterns, making it a tiring task for the practitioner. 
A scenario that is further complicated in the Global South. Indeed, epilepsy imposes a significant health burden in particularly in low- and middle-income countries (LMICs), where over 80\% of the world's 50 million people with epilepsy reside. These regions experience higher prevalence rates due to factors such as perinatal complications, traumatic brain injuries, and infections of the central nervous system 
\cite{chin2012epilepsy}.  
Misconceptions about epilepsy also lead to social stigma, especially in Sub-Saharan Africa, discouraging individuals from seeking medical help and resulting in discrimination in education and employment \cite{radhakrishnan2009challenges}. 
Lastly, the lack of treatment increases the risk of premature death, with individuals in LMICs facing up to three times higher mortality rates compared to those in high-income countries. Furthermore, about 13 million disability-adjusted life years (DALYs) are lost each year as a result of untreated epilepsy, which significantly increases impairment. Recent estimates point out that this burden is mostly from the Global South \cite{singh2020global}. 
 
Over the past decade, substantial progress has been made in the field of artificial intelligence (AI), and a range of tools have been proposed, including those for epilepsy.  For instance, an algorithm was developed to differentiate between idiopathic generalized and focal epilepsy in individuals aged 10 years and older. This model achieved a precision and sensitivity of 0.81, and a specificity of 0.77 
\cite{asadi2023epilepsy}. 

Despite these advancements, the standardization of EEG data, the generalizability of the model among diverse populations, and the integration of AI tools into routine clinical workflows remain challenges. In addition, most of these outstanding results are achieved with high-standard clinical machines in controlled settings such as hospitals. 

In this paper, we employ attention-based graph neural networks (GNNs) to process EEG signals represented as graphs, enabling automatic detection of epileptic brain signals—even in the absence of seizures. Our data were acquired in rural areas using affordable EEG devices, and AI analysis was conducted with limited computational resources. Unlike prior studies using graph convolutional networks (GCNs) or graph attention networks (GATs) to classify ictal vs. pre-ictal states 
\cite{wang2022spatiotemporal, mazurek2024explainable}, our work focuses on differentiating resting-state EEG between epileptic patients and controls, using consumer-grade hardware.
Moreover, those studies have used high resolution data coming from clinical devices, while our analysis has also been carried out voluntarily with limited resources.

The overall pipeline is shown in Figure   \ref{fig:pipeline}: 
 First, we perform the pre-processing of the signals and extract the descriptive features followed by 
 finding the hyperparameters of the network, all carried out with accessible cloud computing and tested on cost-effective computers like a Raspberry PI. Finally, we analyze the results in terms of feature relevance for predicting a given class and exploit the learned attention coefficients to examine connectivity changes between the subject's brain regions, and 
we show that the clinical connections identified by the network are consistent with those observed by practitioners in the literature.  

Given two populations represented using graphs, an effective approach to identifying discriminative characteristics is the network-based statistic (NBS) \cite{zalesky2010network}, which is a statistical method designed to detect significant differences in brain connectivity networks between groups, such as patients and healthy controls, as in our case. 
This tool performs edge-wise comparisons by using a combination of t-tests,  a multiple comparisons correction by identifying clusters of interconnected supra-threshold edges, and assessing their statistical significance through permutation testing. 
Nevertheless, attention mechanisms can learn which nodes or edges are most important for a predictive task, potentially capturing complex nonlinear dependencies that NBS might miss. Moreover, following previous frameworks based on multilink and linear discriminant analysis \cite{crimi2019multilink},  
attention mechanism can be part of an end-to-end trainable model, allowing biomarkers to emerge naturally during training on large datasets.

 Ultimately, although explainability is our main aim rather than classification, the classification results are compared to a random forest classifier and a simpler graph convolutional network (GCN) without attention. 
The choice of random forest classifier rather than other possible machine learning algorithms is given by the consistency with the seminal work with this dataset \cite{van2018reliable}, and the investigation with GCN is related to the question whether the attention mechanism has an impact apart from the explainability in graph based classifier. 

\section{Methods}

GNNs have the ability to learn complex  representations by leveraging the underlying topological relationships in data \cite{velivckovic2023everything,wu2019graphsurvey}, providing   a non-Euclidean framework   \cite{BronsteinGeometricDL}. Formally, a graph is defined as $\mathcal{G} = (\mathcal{V}, \mathcal{E})$, where $\mathcal{V} = \{v_1, \ldots, v_n\}$ represents the set of $n$ nodes, and $\mathcal{E} = \{(v_i, v_j) \mid v_i, v_j \in \mathcal{V}\}$ denotes the set of edges, each indicating a relationship between node $v_i$ and node $v_j$. Usually, each node $v_i \in \mathcal{V}$ is also associated with a feature vector $\mathbf{x}_{v_i}^{(0)} \in \mathbb{R}^{d^{(0)}}$, represented as a column vector.

\subsection{Graph Attention}
In typical GNN architectures, each layer $l$ updates the hidden state of a node by aggregating information from its neighbors. This process can generally be described as:
\begin{equation}
\mathbf{x}_{v_i}^{(l+1)} = \phi \left( \mathbf{x}_{v_i}^{(l)}, \mathcal{A} \left( \{ \mathbf{x}^{(l)}_{v_j} \mid v_j \in \mathcal{N}_{v_i} \} \right) \right),
\end{equation}
where $\mathbf{x}_{v_i}^{(l+1)} \in \mathbb{R}^{d^{(l+1)}}$ is the updated representation of node $v_i$, $\mathcal{N}_{v_i}$ denotes the set of neighbors of $v_i$, $\phi$ represents a non-linear activation function, and $\mathcal{A}$ is a permutation-invariant aggregation function. Various aggregation strategies have been proposed in the literature \cite{velivckovic2023everything}. Among these, attention-based mechanisms offer a compelling approach by including the importance of learned edges during neighborhood aggregation \cite{velickovic2018graph}. Following the approach in \cite{brody2021gatv2}, and using computationally  the GATv2Conv layer, we calculate an attention score $s_{v_i, v_j}^{(l)}$ for each edge $(v_i, v_j)$:
\begin{equation}
s_{v_i, v_j}^{(l)} = \mathbf{a}^{(l)\top} \cdot \text{LeakyReLU} \left( \mathbf{W}^{(l)} \cdot [\mathbf{x}_{v_i}^{(l)} || \mathbf{x}_{v_j}^{(l)}] \right),
\end{equation}
where $\mathbf{a}^{(l)} \in \mathbb{R}^{d^{(l+1)}}$ and $\mathbf{W}^{(l)} \in \mathbb{R}^{d^{(l+1)} \times 2d^{(l)}}$ are learnable parameter matrices, respectively, and $||$ denotes the concatenation of the feature vectors $\mathbf{x}_{v_i}^{(l)}$ and $\mathbf{x}_{v_j}^{(l)}$. These scores are then normalized across the neighbors of node $v_i$, including $v_i$ itself:
\begin{equation} \label{score_alpha}
\alpha_{v_i, v_j}^{(l)} = \frac{\exp\left(s_{v_i, v_j}^{(l)}\right)}{\sum_{v_k \in \mathcal{N}_{v_i} \cup \{v_i\}} \exp\left(s_{v_i, v_k}^{(l)}\right)},
\end{equation}
where $\alpha_{v_i, v_j}^{(l)}$ represents the resulting attention coefficient for the edge connecting $v_i$ and $v_j$. Finally, the updated representation for node $v_i$ at layer $l+1$ is computed by a weighted sum of the transformed feature vectors of its neighbors and itself:
\begin{equation}
\mathbf{x}_{v_i}^{(l+1)} = \phi \left( \sum_{v_j \in \mathcal{N}_{v_i} \cup \{v_i\}} \alpha_{v_i, v_j}^{(l)} \mathbf{M}^{(l)} \cdot \mathbf{x}_{v_j}^{(l)} \right),
\end{equation}
where $\mathbf{M}^{(l)} \in \mathbb{R}^{d^{(l+1)} \times d^{(l)}}$ is a learnable transformation matrix. To enhance learning stability \cite{velickovic2018graph,brody2021gatv2}, employing multiple attention \textit{heads} is a common practice. For each head, an independent set of attention scores $\bm{\alpha}^{(l)}_h$ is calculated (corresponding to Eq. (\ref{score_alpha})). The final output of the $l$-th layer for each node is obtained by either concatenating or averaging the outputs from all $N_h^{(l)}$ attention heads. 

\subsection{Explainable Epileptic Functional Connectivity Detection with GNNs}

As mentioned above, the main goal of this work is to focus on the characteristics of the network that are indicative of epileptic traits derived from the EEG data. Our focus is on providing interpretable explanations by identifying the most salient edges and features, which will be elaborated upon later. 

We employ the learned attention coefficients to assess the importance of connections within the graph. Specifically, we average the attention scores $\alpha_{v_i, v_j}^{(l)}$ (from Eq. (\ref{score_alpha})) across all $N_h^{(l)}$ attention heads in the final GATv2Conv layer to quantify edge significance.  
To understand the importance of particular nodes in the graph, we employ the well-known Grad-CAM method \cite{Selvaraju2017GradCAM}.

\subsection{Model Architecture and Training}
Our final model comprises two GATv2Conv layers (6 attention heads, 32 hidden channels, LeakyReLU activation), followed by a softmax layer. Training used the Adam optimizer (learning rate $1 \times 10^{-3}$, weight decay $5.8 \times 10^{-3}$) with leave-one-out cross-validation. Hyperparameters were tuned via grid search. Details of the architecture and the hyper-paramters search are in the supplementary material. 
To ensure accessibility, we trained the model on Google Colab and deployed it on a Raspberry Pi 4 (8\,GB RAM), demonstrating feasibility for low-resource settings. 

The implementation uses Python 3.10, Pytorch 1.13.1 \cite{PaszkeTorch}, Pytorch Geometric 2.2 \cite{FeyPytorchGeometric}, and Nilearn 0.11.1 (\url{https://nilearn.github.io/}) for brain structure visualization.
The code is available at URL 
\url{https://github.com/alecrimi/eeg_nigeria_guinea}.


\begin{figure*}[t!]
  \centering
  \subfloat[][ ]{
    \includegraphics[width=0.32\textwidth]{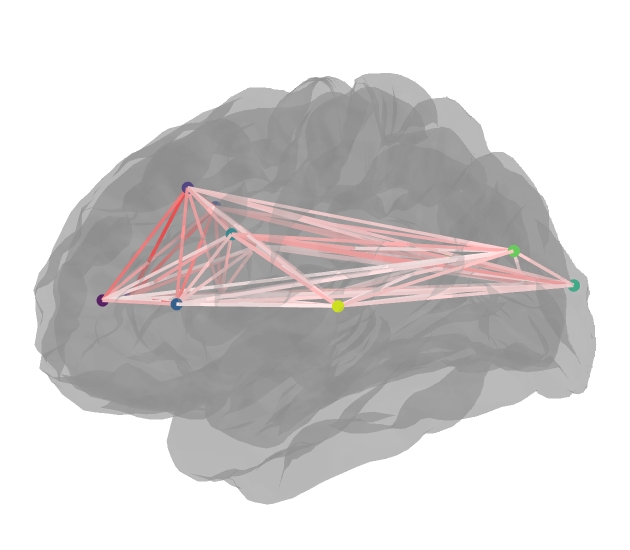}
 
  }
  \hfill
  \subfloat[][ ]{
    \includegraphics[height=150pt,width=0.25\textwidth]{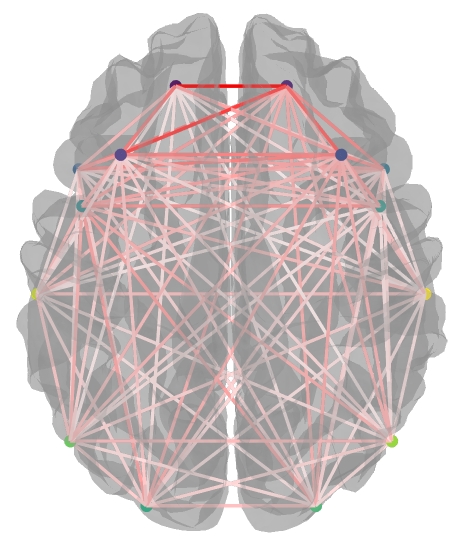}
 
  }
  \hfill
  \subfloat[][ ]{
    \includegraphics[width=0.32\textwidth]{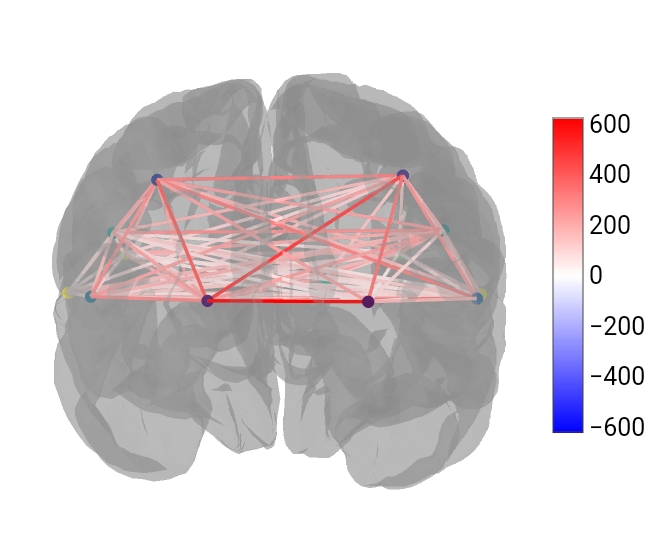}
 
  }

  \caption{Attention weights related to the brain connectivity for the Nigeria dataset: (a) sagittal, (b) axial, and (c) coronal view, showing stronger weights in the fronto-temporal area. For the labels of the nodes, please refer to Figure~1. The color code marks the dark red connections as most relevant as those among AF3, AF4, F3, F4, FC5 and FC6.}
  \label{fig:attention_weights_ni}
\end{figure*}

\section{Data and experimental settings}

\subsection{Dataset and pre-processing}
The used data were acquired from people with epilepsy (N=163) and healthy controls (N=138) in two difficult-to-reach areas in rural Guinea-Bissau and Nigeria, in accordance with international standards including written consents \cite{van2018reliable}, and are publicly available \footnote{\url{https://zenodo.org/records/1252141}}. From the Nigerian dataset 5 subjects were removed due to high signal to noise ratio, as reported in the table of the repository \cite{van2018reliable}.

Five minutes of fourteen-channel resting-state EEG data were acquired with a portable, low-cost consumer-grade EEG recording headset Epoc (Emotiv Inc, CA, USA).  The Epoc headset has 14 EEG channels, plus 2 reference sensors.
The recordings were approximately five minutes	(300 s) at a sampling rate of	128 Hz. Approximately the first three minutes were with the open-eyes protocol, while the last 2 minutes were with the closed-eyes protocol. In the Nigerian dataset some recording were all closed-eyes protocol.

The data were then divided into 5-second epochs with 1-second overlap. Therefore, for each subject 
window $w=1,\ldots,N_{sw}$, and channel $i=1,\ldots,14$ we created a set of time series $r_{swi}(t)$. 
Afterwards, data were processed by going through a pass-band filter (0.5-45 Hz), and independent component analysis-based motion correction using 10 components. The resulting connectivity matrix related to each graph is in the end $14\times 14 = 196$.

\subsection{Graph representation and feature extraction} 

For each sample $r_{swi}(t)$, we represent the relationships between electrodes as a complete graph. We additionally weight the edges between a given pair of nodes by assigning the corresponding phase locking value \cite{bruna2018plv}. This approach has proven useful in other studies utilizing GNNs for epileptic EEG classification, showing an advantage over other graph construction methods\cite{wang2023plvgraphs}. 
Manual extraction of characteristics from raw time series processing in EEG analysis has been proven to be effective in previous studies \cite{JiaEfficientGraphConv, wijayanto2019fdepilepsy}.
With the aim of a lightweighted system potentially running on RaspberryPI or low-end computers, we choose few features to extract, demanding little computations. In practice, for each training and testing sample, we extracted
the Katz fractal dimension \cite{esteller2001katzfd},
and the energy of the delta (0.5-4 Hz), theta (4-8 Hz), alpha (8-12 Hz) and beta (12-29 Hz) bands. These constituted the components of the feature vectors $\textbf{h}_{swi}^{(0)}, \ \forall i \in \mathcal{V}_{sw}$.
An investigation of the impact of specific features and the trade-off with performance and classification is beyond the scope of this work.


\section{Results}

The experiments were conducted in a leave-one-out cross-validation manner. The classification results for the Nigerian and Guinea datasets are reported for the random forest, GCN,  and GAT classifiers respectively in Table \ref{tab:nigeria_classification_comparison} 
and \ref{tab:gb_classification_comparison}.

\begin{figure*}[htbp]
  \centering
  \subfloat[][ ]{
    \includegraphics[width=0.32\textwidth]{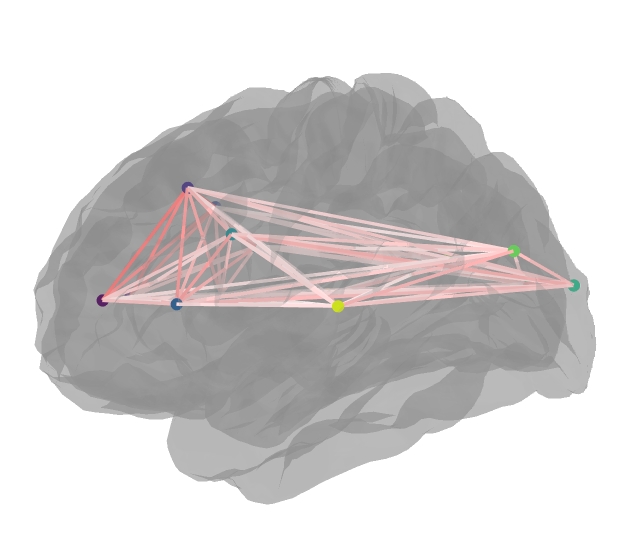}
 
  }
  \hfill
  \subfloat[][ ]{
    \includegraphics[height=150pt,width=0.25\textwidth]{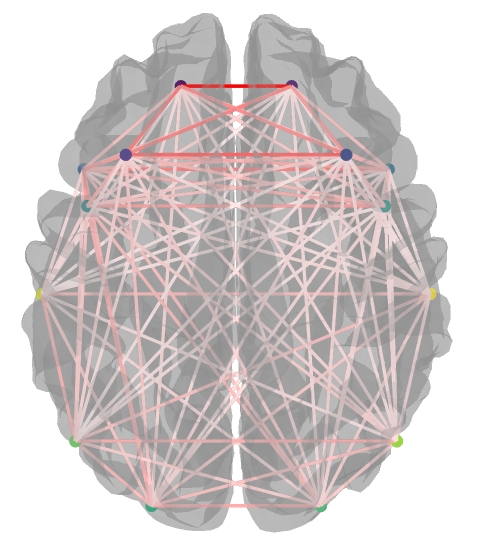}
 
  }
  \hfill
  \subfloat[][ ]{
    \includegraphics[width=0.32\textwidth]{coronal_ni.jpg}
 
  }

  \caption{Attention weights related to the brain connectivity for the Guinea-Bissau dataset: (a) sagittal, (b) axial, and (c) coronal view, showing stronger weights in the fronto-temporal area. For the labels of the nodes, please refer to Figure~1. The color code marks the dark red connections as most relevant as those among AF3, AF4, F3, F4, FC5 and FC6.}
  \label{fig:attention_weights_gb}
\end{figure*}

\begin{figure}[htbp]
  \centering
  \includegraphics[width= \columnwidth]{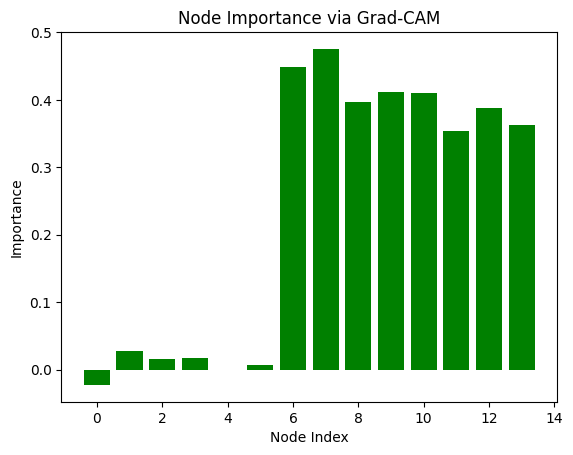}
  \caption{Attention weights at node level for the Nigerian dataset computed by Grad-CAM method. Here the nodes with highest weights are the last listed, which are for example AF3, AF4, F3, F4, F7, F8, FC5 and FC6.}
  \label{fig:node_ni}
\end{figure}

\begin{figure}[htbp]
  \centering
  \includegraphics[width= \columnwidth]{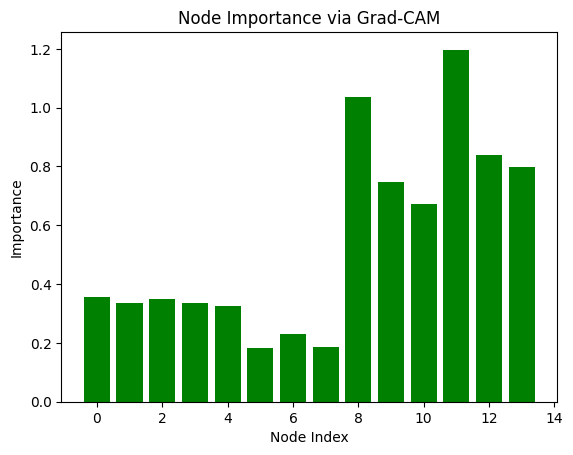}
  \caption{Attention weights at node level for the Guinea-Bissau dataset computed by Grad-CAM method. Here the nodes with highest weights are the last listed, which are for example AF3, AF4, F3, F4, F7, F8, FC5 and FC6.}
  \label{fig:node_gb}
\end{figure}

Moreover, to validate that differences between classifier performances are significant, we conducted paired statistical tests by using the  DeLong test, comparing AUROC values. We obtained  p-values $<0.05$ for both datasets and between all classifiers. 
The resulting attention weights for Nigeria and Guinea datasets are reported in Figures \ref{fig:attention_weights_ni} and \ref{fig:attention_weights_gb} respectively, while Figure   \ref{fig:node_ni} and  \ref{fig:node_gb} report the node importance scores. 
During the architecture search, we observe that in GAT layers, going beyond 6 attention heads and 32 hidden channels was not improving the results, while increasing the computational costs. Using 2 heads, or fewer than 16 hidden channels was leading to performance deterioration.


The entire pipeline of preprocessing and training the model for one dataset in a cross-validation setting took approximately 30 minutes per dataset on a free Google Colab environment. 

\begin{table*}[ht]
\centering
\caption{Nigeria Dataset Classification Report Comparison (Random Forest, GCN, GAT).}
\begin{tabular}{lcccccc}
\toprule
\textbf{Metric} & \multicolumn{2}{c}{\textbf{Random Forest}} & \multicolumn{2}{c}{\textbf{GCN}} & \multicolumn{2}{c}{\textbf{GAT}} \\
\cmidrule(r){2-3} \cmidrule(r){4-5} \cmidrule(r){6-7}
 & Control & Epilepsy & Control & Epilepsy & Control & Epilepsy \\
\midrule
Precision      & 0.72 & 0.67 & 0.55 & 0.64 & 0.67 & 0.72 \\
Recall         & 0.77 & 0.62 & 0.46 & 0.72 & 0.60 & 0.77 \\
F1-score       & 0.74 & 0.65 & 0.50 & 0.67 & 0.64 & 0.75 \\
\midrule
Accuracy       & \multicolumn{2}{c}{0.70} & \multicolumn{2}{c}{0.61} & \multicolumn{2}{c}{0.70} \\
Macro avg F1   & \multicolumn{2}{c}{0.70} & \multicolumn{2}{c}{0.59} & \multicolumn{2}{c}{0.69} \\
Weighted avg F1& \multicolumn{2}{c}{0.70} & \multicolumn{2}{c}{0.60} & \multicolumn{2}{c}{0.70} \\
AUROC          & \multicolumn{2}{c}{\textbf{0.70}} & \multicolumn{2}{c}{\textbf{0.64}} & \multicolumn{2}{c}{\textbf{0.78}} \\
\bottomrule
\end{tabular}
\label{tab:nigeria_classification_comparison}
\end{table*}

 \begin{table*}[ht]
\centering
\caption{Guinea-Bissau Dataset Classification Report Comparison (Random Forest, GCN, GAT).}
\begin{tabular}{lcccccc}
\toprule
\textbf{Metric} & \multicolumn{2}{c}{\textbf{Random Forest}} & \multicolumn{2}{c}{\textbf{GCN}} & \multicolumn{2}{c}{\textbf{GAT}} \\
\cmidrule(r){2-3} \cmidrule(r){4-5} \cmidrule(r){6-7}
 & Control & Epilepsy & Control & Epilepsy & Control & Epilepsy \\
\midrule
Precision      & 0.66 & 0.73 & 0.65 & 0.72 & 0.69 & 0.76 \\
Recall         & 0.79 & 0.59 & 0.72 & 0.55 & 0.75 & 0.71 \\
F1-score       & 0.72 & 0.66 & 0.72 & 0.61 & 0.75 & 0.71 \\
\midrule
Accuracy       & \multicolumn{2}{c}{0.69} & \multicolumn{2}{c}{0.61} & \multicolumn{2}{c}{0.73} \\
Macro avg F1   & \multicolumn{2}{c}{0.69} & \multicolumn{2}{c}{0.59} & \multicolumn{2}{c}{0.73} \\
Weighted avg F1& \multicolumn{2}{c}{0.69} & \multicolumn{2}{c}{0.60} & \multicolumn{2}{c}{0.73} \\
AUROC          & \multicolumn{2}{c}{\textbf{0.76}} & \multicolumn{2}{c}{\textbf{0.70}} & \multicolumn{2}{c}{\textbf{0.82}} \\
\bottomrule
\end{tabular}
\label{tab:gb_classification_comparison}
\end{table*}

\section{Discussions}
The proposed graph attention network outperformed the random forest classifier, as shown comparing the tables especially for the AUROC metric. 
There is a visible difference in classification between countries, with the Nigerian dataset being classified less accurately compared to the Guinea-Bissau dataset.  This is less pronounced using the GAT classifier, though still present. This has also been observed in the original research \cite{van2018reliable}, which is puzzling, since the same device was used and almost the same specialists acquired the data. It is difficult to speculate on a difference in the populations. It can be hypothesized that the difference could be related to the slightly different protocol, as Nigerian EEG acquisitions were performed with the first 3 minutes with the eyes open and then 2 minutes with the eyes closed, rather than further separating the two types of acquisition. Additionally, EEG activity characteristics vary greatly between individuals, therefore such differences are anticipated unless the patient cohort is very large. 
While it is true that AUROC was always better for GAT, random forest occasionally showed slightly better performance in individual metrics like F1-score or recall, our choice of AUROC as the primary metric is deliberate. The AUROC is threshold independent, robust to class imbalance, and better captures the overall discriminative power of the model, which is especially important in medical screening scenarios like epilepsy detection, and we consider it more reliable given the small sample size of the data. Despite both being graph neural networks, GCN did not perform well as the GAT classifier. This might be related to the fact that GCNs treat all neighbors equally during message passing in case of unweighted graphs. In EEG, not all electrodes (nodes) contribute equally to the seizure pattern; some may carry more critical signals. GCN cannot effectively weigh this difference. On the other hand, GAT introduces attention mechanisms that assign different weights to neighbors. This helps capture heterogeneity between brain regions, which is especially important in epilepsy detection, where some areas are more epileptogenic.
Another explanation is that constructed functional connectivity is prone to noise, and GCNs cannot learn meaningful representations, while GATs can still adapt, by ignoring irrelevant connections through attention weights. Nevertheless, this highlights that the case in exam is not a trivial problem. It is possible that other legacy classifiers  - as random forest - achieve high classification, but they might behave as black-box and require integration of other explainability tools. In this view, we see GAT as achieving acceptable classification results and at the same time allowing visualization of the attention coefficients on the edges. 

Even if the classification results for the two datasets were different, from the point of view of attention features, the results were consistent across the datasets.  
Indeed, analyzing the attention weights of graph features reveals that the nodes most relevant for classification are AF3, AF4, F3, F4, FC5, and FC6. These nodes correspond to the frontal, frontotemporal, and temporal brain areas.
This is consistent with the inspection of relevant connections. In fact, the connections with the highest weights in Figures \ref{fig:attention_weights_ni} and \ref{fig:attention_weights_gb}, for both datasets, the connections within the frontal area, and the connections between the frontal and temporal areas are the most relevant, although some frontal-occipital connections are also present.
For example, the connection between AF3 and AF4, and among F3, F4, F7, F8, FC5, and FC6, also some between F3/F4 and T7/T8/01/02.  Interestingly, despite the difference in classification depending on the dataset, no significant differences were observed in the edges highlighted by the attention weights. The fact that attention weights may be more focused on the frontal areas or frontotemporal regions is due to the frontal and temporal lobes playing an important role in conditions like epilepsy. 
In fact, epileptic seizures often involve the frontal lobes, as they are responsible for controlling motor functions and higher cognitive processes. Many seizures, especially focal  seizures, originate in these areas, especially around the motor cortex, prefrontal cortex, and insula \cite{mcgonigal2022frontal}. The frontotemporal network is also involved in maintaining focus and cognitive control, and disruptions in this area are often observed in epilepsy. 
The temporal lobes are also frequently implicated, especially in temporal lobe epilepsy, which is one of the most common forms of epilepsy \cite{centeno2014network}.  
In some cases, the frontotemporal areas show decreased connectivity due to seizure activity, while other studies show increased synchronization in these areas during interictal periods (between seizures) \cite{laufs2014altered}.


Several studies have proposed graph-theoretical network analysis to study complex networks arising from the structural connections of epileptic patients. 
These studies focused primarily on the search for connectivity alterations using graph metrics \cite{lin2020altered}.  Looking at the studies using GATs \cite{wang2022spatiotemporal,mazurek2024explainable},   using high-end clinical devices, they ended up using similar GAT architectures, though obtaining better performances with 128 hidden layers, while for our experiments, there was no significant improvement. 
Future works include collection of further datasets in other countries
, and the implementation of federated learning.

\section{Conclusions}
Despite the fact that epilepsy is a complicated topic with many variations and many heterogeneous causes, AI models can be easily used to classify EEG data for diagnosing epilepsy, and not just for detecting seizures. 
This is particularly feasible for low-income countries where portable EEG devices can be used without losing quality, and model training can be performed via online platforms such as Google Colab. 





\bibliographystyle{IEEEtran}
\bibliography{OJEMBTemplate}

 \clearpage 

\section*{Supplementary Material}

\subsection{Details of Model Architecture and Training}
We explored many network architectures, and the final model comprised two graph attention layers (corresponding to $l=0$ in the preceding equations), with $N_h^{(0)} = 6$ attention heads and a LeakyReLU activation function ($\phi(\cdot) \equiv \text{LeakyReLU}(\cdot)$), and 32 hidden feature channels. 
A final softmax layer produced the probability distribution for the three output classes.

More precisely, the model takes the following inputs:
\begin{itemize}
    \item Node feature matrix $X \in \mathbb{R}^{N \times F}$, where $N$ is the number of nodes and $F = 5$ is the number of features per node.
    \item Edge index $E$, defining the graph's connectivity in coordinate format.
    \item Edge attributes $A$, representing features associated with each edge.
    \item A batch vector $B$, indicating the graph affiliation of each node for mini-batch processing.
\end{itemize}

The core of the architecture is the GATv2Conv module. The initial layer is a multi-head attention mechanism with $h$ attention heads, projecting the input node features into a $d$-dimensional hidden space. Each head independently computes attention coefficients across the neighboring nodes. The resulting outputs are then concatenated:
\begin{equation}
H^{(1)} = \text{ELU}(\text{LayerNorm}(\text{GATv2Conv}(X, E, A)))
\end{equation}
where $H^{(1)} \in \mathbb{R}^{N \times (h \cdot d)}$.

This is followed by a second GATv2Conv layer organized similarly. This layer refines the node embeddings while maintaining their dimensionality. Layer normalization and an ELU activation function are applied:
\begin{equation}
H^{(2)} = \text{ELU}(\text{LayerNorm}(\text{GATv2Conv}(H^{(1)}, E, A)))
\end{equation}

After obtaining the final node embeddings $H^{(2)}$, a global add-pooling operation aggregates the node-level information to produce a graph-level embedding:
\begin{equation}
Z = \text{GlobalAddPool}(H^{(2)}, B)
\end{equation}

This graph-level representation $Z$ is then processed by a multi-layer perceptron consisting of two fully connected layers. An intermediate batch normalization layer and a ReLU activation function are applied after the first linear layer:
\begin{equation}
Z' = \text{ReLU}(\text{BatchNorm}(\text{Linear}_1(Z)))
\end{equation}
\begin{equation}
\text{Output} = \text{LogSoftmax}(\text{Linear}_2(Z'))
\end{equation}

The final output is a log-probability distribution over the target classes. In this specific configuration, the model is designed for binary classification ($\text{out\_channels} = 2$).

This architecture is supposed to be well-suited for scenarios where the relationships between nodes are critical, as anticipated in (EEG) brain graph classification. The incorporation of edge attributes and layer normalization contributes to the model's stability and expressive power.

The model was trained using the Adam optimizer with a learning rate of $1 \times 10^{-3}$ and a weight decay of $5.8 \times 10^{-3}$, for a total of 400 training iterations. Evaluation was performed using a leave-one-out cross-validation strategy. Performance for each fold was assessed using the Area Under the Receiver Operating Characteristic Curve (AUROC), F1-score, precision, recall, and precision. An early stopping mechanism was implemented to halt training if no improvement in the validation loss was observed for 10 consecutive epochs, although this condition was not met during training. A dynamic learning rate scheduler was also investigated, but did not enhance performance. Given the leave-one-out cross-validation, the reported metrics are averaged across all training folds.

The hyperparameters have been tuned through a grid search using stratified cross-validation in a non-exhaustive manner. The search space included the following ranges: the number of GAT layers was varied between 1 and 3; the number of hidden units per layer was set to one of \{8, 16, 32, 64\}; the number of attention heads was selected from \{2, 4, 6, 8\}; dropout rates considered were 0.0, 0.2, and 0.5. For optimization, we experimented with learning rates of 1e-2, 5e-3, and 1e-3, and weight decay values of 0, 1e-4, and 5e-4, 5.8e-3. In the Result section we report ablation observations as well as observation on extreme values of these intervals. 
Initially, the GCN classifier to compare was tested using the final optimal GAT architecture. However, this gave very poor results, therefore an indipendent grid search was applied with the same ranges to the GCN classifier but adding also 2 further layers. The resulting architecture was therefore composed by a sequence of four graph convolutional layers. After each convolutional layer, the model applied batch normalization, which helps stabilize the learning process by normalizing feature distributions across the batch. This is followed by a ReLU activation. Finally, a log-softmax function is applied to produce class probabilities in log-space.

\end{document}